\documentclass[12pt]{iopart}
\usepackage{graphicx}

\def\v#1{{\bf#1}}
\def\bea{\begin{eqnarray}}
\def\eea{\end{eqnarray}}
\def\ahalf{{\textstyle{1\over2}}}
\newcommand{\bfalpha}{\mbox{\boldmath$\alpha$\unboldmath}}

\begin{document}
\title[Schematic baryon models ...]{Schematic baryon models, their tight binding description and their microwave realization}
\author{E. Sadurn\'i$^1$, J. A. Franco-Villafa\~ne$^2$, U. Kuhl$^3$, F. Mortessagne$^3$, and T. H. Seligman$^{2,4}$}

\address{$^1$ Instituto de F\'isica, Benem\'erita Universidad Aut\'onoma de Puebla,
Apartado Postal J-48, 72570 Puebla, M\'exico}

\address{$^2$ Instituto de Ciencias F\'isicas, Universidad Nacional Aut\'onoma de M\'exico, Av.
Universidad s/n, 62210 Morelos, M\'exico.}

\address{$^3$ Universit\'e Nice Sophia Antipolis, CNRS, Laboratoire de Physique de la Mati\`ere Condens\'ee, UMR 7336 Parc Valrose, 06100 Nice, France.}

\address{$^4$ Centro Internacional de Ciencias, Universidad Nacional Aut\'onoma de M\'exico, Av.
Universidad s/n, 62210 Morelos, M\'exico.}

\eads{\mailto{sadurni@ifuap.buap.mx}}
\submitto{\NJP}

\begin{abstract}
A schematic model for baryon excitations is presented in terms of a symmetric Dirac gyroscope, a relativistic model solvable in closed form, that reduces to a rotor in the non-relativistic limit. The model is then mapped on a nearest neighbour tight binding model. In its simplest one-dimensional form this model yields a finite equidistant spectrum. This is experimentally implemented as a chain of dielectric resonators under conditions where their coupling is evanescent and a good agreement with the prediction is achieved.
\end{abstract}

\maketitle

\section{Introduction}
\label{sec:Intro}

In the sequel of the surge of interest in graphene \cite{nov05} and its connection to the Dirac equation, emulations of relativistic equations by analogue systems and their experimental realization have been boosted. There are several realization of artificial graphene \cite{pol13,rei13}, i.e., a honeycomb lattice structure, like microwave systems \cite{pel07,bit10a,zan10,bit12,kuh10a,bel13a,bar13a,bel13b}, molecular graphene \cite{gom12} or ultracold atoms in optical lattices \cite{tar07}. But not only honeycomb lattices have been realized, also one-dimensional systems, where Klein-Tunneling was observed \cite{dre12} or the Dirac-Moshinsky oscillator\cite{sad10a} was realized \cite{arXfra13}. All those experiments show the interest to investigate relativistic systems in general and realize them in analogue experiments. The relation is based on the symmetry itself which yields the well known Dirac points \cite{sem80}. Nearest neighbour interaction hamiltonians have been used for a long time and are at the base of many of these models, although in practice both for graphene and most models higher interaction terms complicate the picture to some extent, as was established quite early in \cite{wal47}.

It now seems of interest to find simple covariant models $-$say for particles$-$ that can be realized in classical wave systems, e.g. microwave experiments. Indeed one of the areas of active research in high energy physics is the investigation of the mass spectrum of baryons starting from quantum chromodynamics in a non-perturbative regime\cite{wil74,eic78}. The task is not a trivial one, as the efforts to obtain answers in this problem are mainly numerical \cite{deg06}. On the other hand, exactly solvable models were proposed with relative success from the very beginning: Attempts in this direction include multi-particle systems with relativistic hamiltonians \cite{isg78,isg79a,isg79b,cha81,cap86}, solvable hamiltonians from a spectrum generating algebra \cite{bij94,bij00} and many-particle Dirac-Moshinsky oscillators \cite{mos90a,mos91,mos92,mos96}.

It has not been easy to obtain these models from first principles. This is in particular true for models involving many quarks, either with a fixed or variable number of them. Despite of some conceptual difficulties, such models have had certain phenomenological success and it is desirable to improve our understanding of even the simplest of them. Therefore it is not a useless endeavour to propose similarly simple constructions, but which actually abandon the multiparticle approach and focus more in structural parameters of hadrons such as size, internal spin and moments of inertia. More clues on the necessity of such models are provided by previous attempts to introduce relativistic oscillators or the more precise `Cornell' potentials between two quarks; their spectrum as a function of the orbital angular momentum $l$ and frozen radial motion is roughly $\sqrt{a l + b}$, where $a$ and $b$ are constants. Since the spectrum we are dealing with is not a concave function\footnote{One can be convinced of this statement by plotting the data published by the PDG \cite{nak10} as a function of $l$. See also our \fref{fig3}.} of $l$, it appears more sensible to introduce a law of the type $\sqrt{a l^2 + b}$.

The general form of this energy suggests a model hamiltonian which resembles the square root of a non-relativistic rotor. We shall see here that such a system can be represented as a tight-binding model with nearest neighbour interaction only. The model can in turn be realized as a one-dimensional array of resonators, thus fulfilling the program of emulating a covariant Dirac-like equation by a microwave experiment. It has furthermore the attraction of displaying a {\it finite} spectrum, and thus it can be realized on a {\it finite} array of resonators. We thus need not worry about cut-off effects, and therefore can focus entirely on questions of reducing systematic and statistical deviations as well as on minimizing second neighbour interactions.

We start by presenting a simple comparison of two baryon excitation spectra with those of the solvable symmetric version of a Dirac gyroscope in \sref{sec:SimplGyro}. In the same section a further simplification is presented which leads to a finite equidistant spectrum. A more complete view and a broader scope is offered in \sref{sec:Zeeman}; this section is not essential to what follows and can be read separately. \Sref{sec:TightBind} discusses the tight binding hamiltonian describing such systems as a one-dimensional array of resonators with nearest neighbour interactions only. These resonators are realized as dielectric disks between two metallic plates in a microwave experiment. We fix parameters on an equidistant chain. The distance of the disks for the realization can be calculated from the corresponding relativistic equations and the coupling strengths is obtained from the experiments with equidistant disks. We do get good agreement between experiment and theory. Finally we proceed to a discussion of these results and possible extensions.

\section{A simplified Dirac gyroscope}
\label{sec:SimplGyro}
\subsection{The gyroscope revisited}
\label{ssec:GyroRevisited}

Based on three postulates for rigidity in a relativistic context \cite{boh83, ald83, ald84, ald85} given below, one of the authors has studied \cite{sad09} a relativistic quantum rotor denominated Dirac gyroscope. It generalizes the Dirac equation to a particle with internal structure. The postulates for such a Poincar\'e invariant formulation are
\begin{itemize}
\item[1.] Elementary limit (standard Dirac equation), reached when the dimensions of the extended object collapse to zero.
\item[2.] Consistent classical limit, recovering a classical relativistic equation.
\item[3.] Consistent non-relativistic limit, reducing the system to a non-relativistic quantum rotor.
\end{itemize}
In the body-fixed frame of reference \footnote{Despite its unusual form, the hamiltonian (\ref{original}) emerges from a Poincar\'e invariant equation containing explicitly the center of mass four-vector and the Pauli-Lubanski vector. A Schr\"odinger-like equation can be obtained when the center of mass is frozen and the inertia tensor is diagonal.}, the corresponding hamiltonian is given by \cite{sad09}

\bea
H= \sqrt{M} c \bfalpha \cdot \left( \bar I \v L \right) + \beta Mc^2,
\label{original}
\eea
where $M$ is the rest mass of the body, $\bfalpha, \beta$ are Dirac matrices in the laboratory frame, $\bar I = {\rm diag\ }\left( I_{xx}^{-1/2}, I_{yy}^{-1/2}, I_{zz}^{-1/2}\right)$ is the inverse square root of the inertia tensor and $\v L$ is the orbital angular momentum vector in the body frame. In equation (24) of \cite{sad09} it was also shown that the corresponding Dirac equation in the body frame leads to a tractable eigenvalue problem when the body has axial symmetry. In the fully asymmetric case, on the other hand, one has to rely on matrix diagonalization methods. For a general tensor of inertia, the spectrum can be rich enough to accommodate levels in various forms. Certain values of the moments of inertia may even yield levels which {\it decrease\ } with an increasing orbital angular momentum number $l$.

\begin{figure}
\begin{center}
\begin{tabular}{cc}
\includegraphics[scale=0.6]{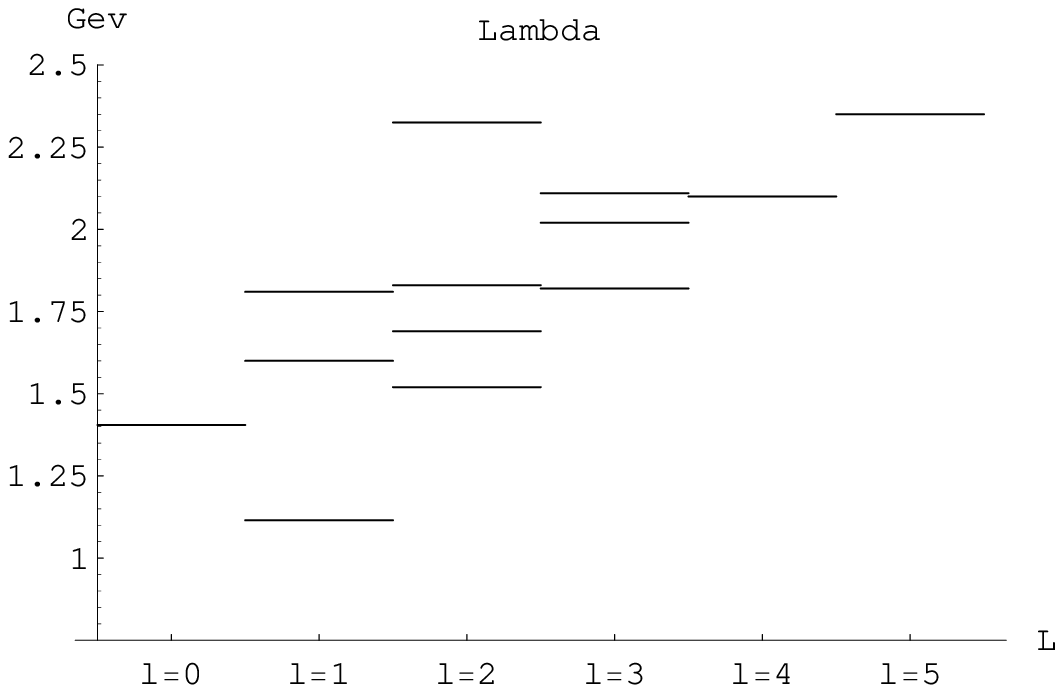}
 & \includegraphics[scale=0.6]{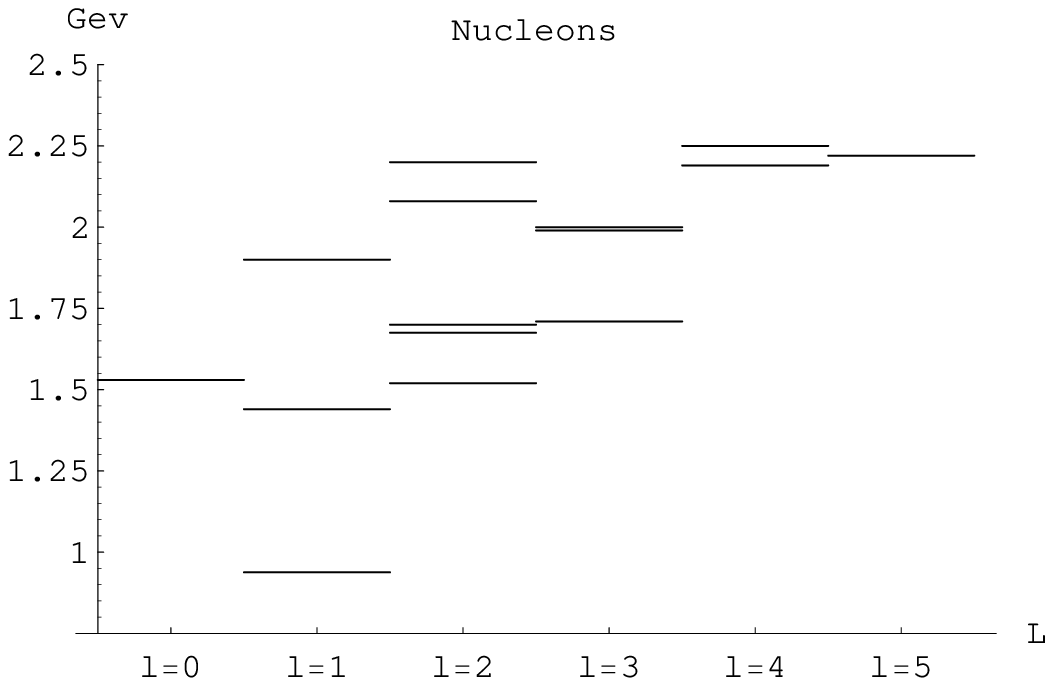} \\
\includegraphics[scale=0.6]{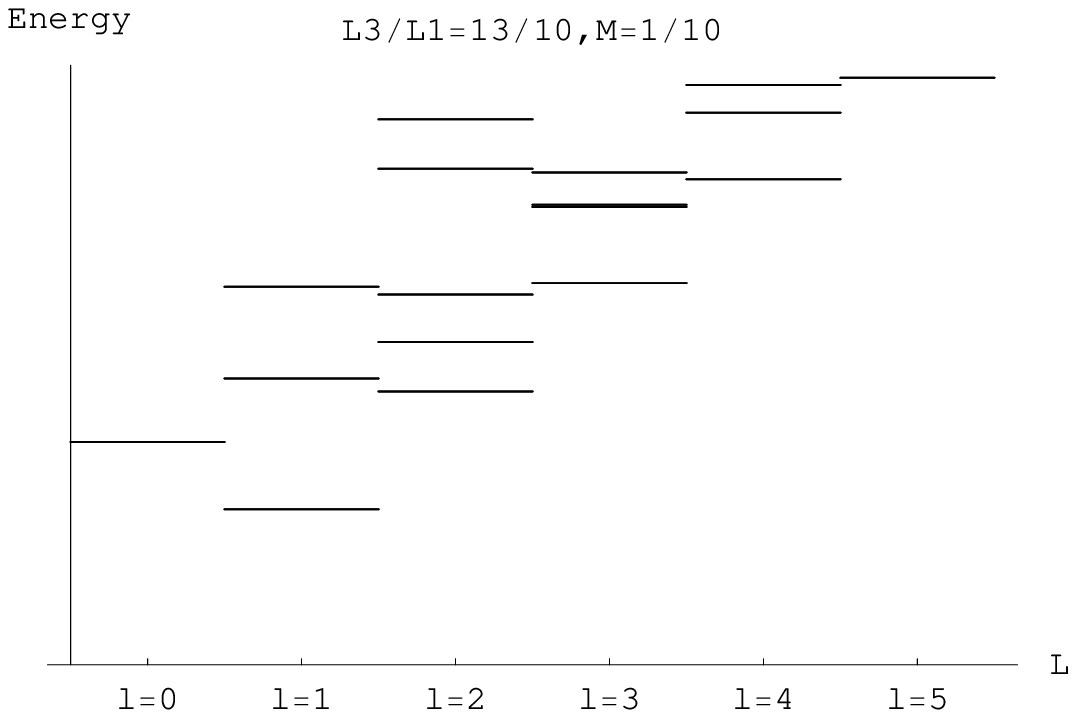}
 & \includegraphics[scale=0.6]{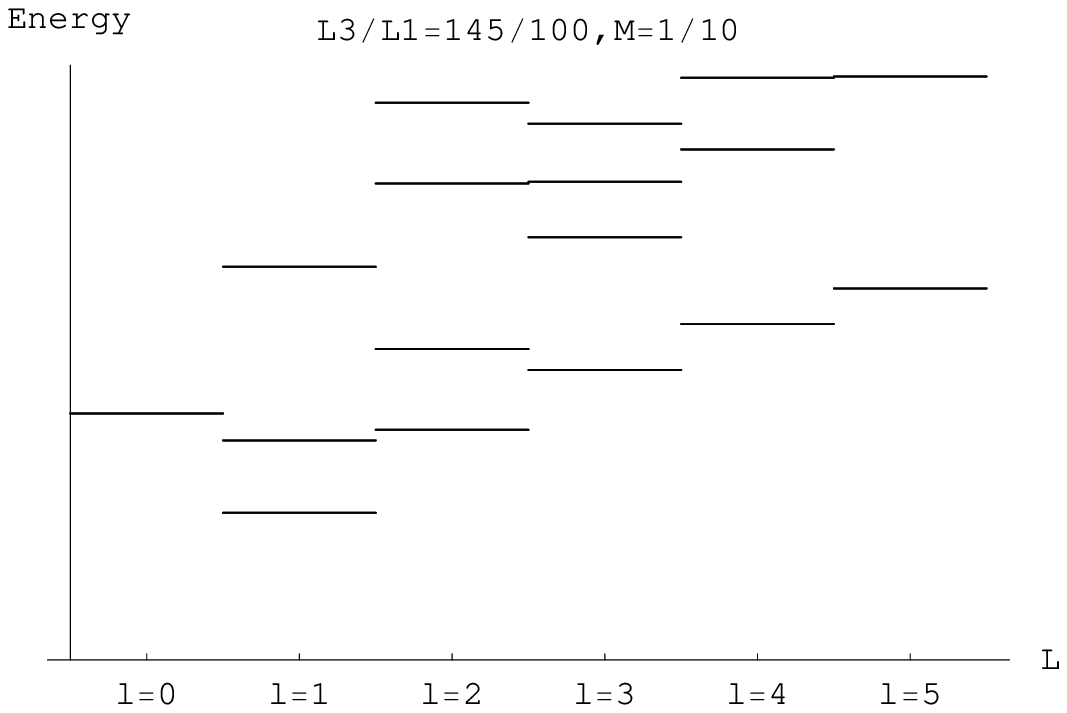}
\end{tabular}
\end{center}
\caption{Upper row: Experimental masses of lambda baryons (quark content uds) and nucleons (quark content udd and uud). Lower row: Theoretical levels using only three parameters in a Dirac gyroscope with axial symmetry. For nucleons we choose $M=1\mbox{GeV}$, $I_{xx}/I = \sqrt{13/10}$ and $I M c^2 / \hbar^2 = 1/10$. For lambda baryons we use $I_{xx}/I = \sqrt{145/100}$. The ratios $I_{xx}/I$ for both cases suggest oblate shapes. Note that, in agreement with experiment, for the lambda particles one $l=1$ level lies below the $l=0$ level, while for nucleons two $l=1$ levels and one $l=2$ level have this property.}
\label{fig3}
\end{figure}

Even for the exactly solvable symmetric case, we find rich spectra. We illustrate this in \fref{fig3}, where we compare two particular spectra of such a gyroscope to mass spectra of Nucleons (udd, uud) and Lambda particles (uds)\cite{nak10}. Indulging the crude identification of quantum numbers, one finds a surprising similarity for low lying levels between a model with axial symmetry and the known data. The simultaneous fit for both spectra can be done by adjusting four parameters. According to the Particle Data Group (PDG) \cite{nak10}, the observables $\sqrt{<R_n^2>}$ (neutron mean square charge radius) and $R_p$ (proton charge radius) have the values $R_p= 0.8 \times 10^{-15}\mbox{m}$, $\sqrt{<R_n^2>} = 0.4 \times 10^{-15}\mbox{m}$. Using the estimate $\hbar c \sqrt{M/I} = \hbar c/ \lambda \sim 1\mbox{GeV}$ for a typical length $\lambda$ of the body, we get $\lambda \sim 10^{-15}\mbox{m}$, which is in the same order of magnitude as the radius of a nucleon.

In the case of Nucleons we choose $M=1\mbox{GeV}$, $I_{xx}/I = \sqrt{13/10}$ and $I M c^2 / \hbar^2 = 1/10$, while in the case of $\Lambda$ baryons we use $I_{xx}/I = \sqrt{145/100}$. The ratios $I_{xx}/I$ for both cases suggest oblate shapes. We have restricted our calculations to an exactly solvable model, but a better fit with the experimental data can be achieved allowing full asymmetry in the body and thus four parameters. Note though, that certain characteristic level inversions between lower and higher angular momenta are present even in this simple model as can be seen in \fref{fig3}.

\subsection{Simplification of the model to a rotor}
\label{ssec:Rotor}

Our purpose is to emulate this relativistic system in one of its simplest forms. We may specialize to limit cases sacrificing neither the relativistic nor the quantum features of the system.

\begin{figure}
\begin{center}
\vspace{0.5cm}
\begin{tabular}{l@{\hspace*{-0.5cm}}cl@{\hspace{-0.5cm}}c}
(a) & & (b) & \\[-6ex]
& \includegraphics[scale=1]{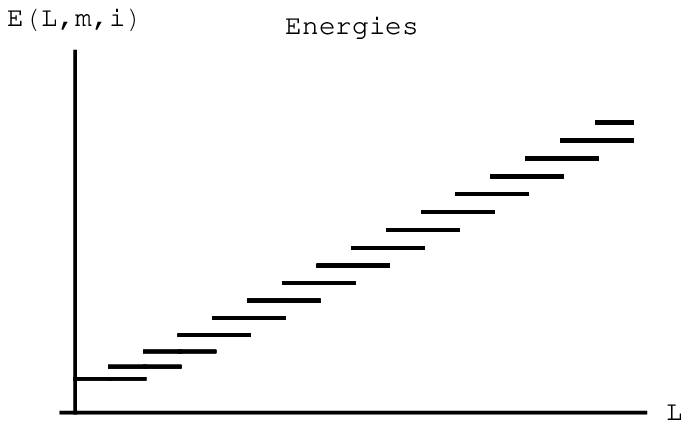}
 & & \includegraphics[scale=1]{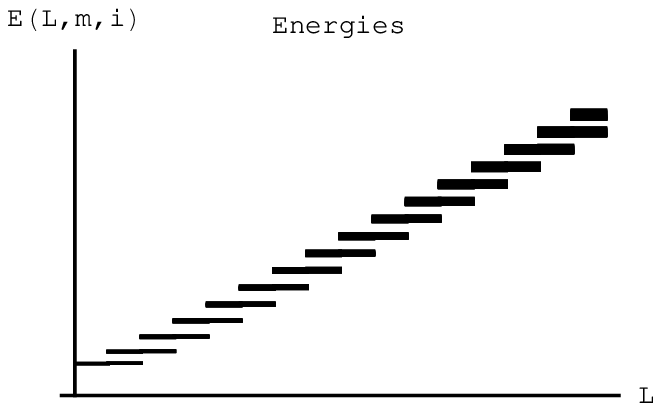} \\[2ex]
(c) & & (d) & \\[-6ex]
& \includegraphics[scale=1]{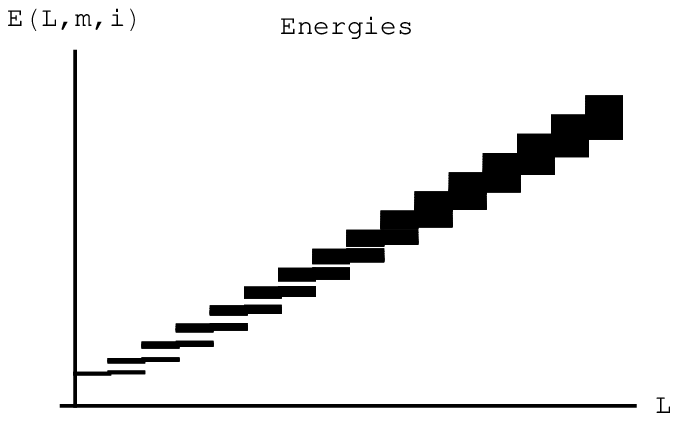}
& & \includegraphics[scale=1]{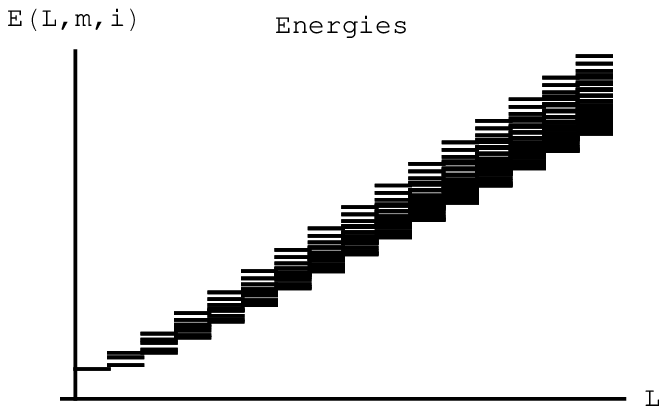}
\end{tabular}
\end{center}
\caption{Energy levels $E(l,m_j)$ of a Dirac gyroscope, shown in a gradual transition from complete symmetry of the inertia tensor to a prolate gyroscope. (a) $I_{xx}/ I = 1$. (b) $I_{xx}/ I = 8/9$. (c) $I_{xx}/ I = 7/9$. (d) $I_{xx}/ I = 2/3$. The levels form bands for each value of $L$. Parameters: $I M c^2 / \hbar^2 = 1/10$.}
\label{fig4}
\end{figure}

For instance, when two elements of the diagonal inertia tensor are large compared to the third - e.g. cigar shaped objects - the terms in the kinetic energy affected by such elements become small and produce a negligible level spacing within certain bands (see \fref{fig4}). It is indeed fair to establish an axial symmetry in this case, since the condition $I_{zz} \approx I_{yy}$ is compatible with the usual restrictions on the moments of inertia

\bea
I_{zz} + I_{xx} > I_{yy}, \qquad I_{yy} + I_{xx} > I_{zz}, \qquad I_{yy} + I_{zz} > I_{xx}
\label{moments}
\eea
as well as with the conditions for a prolate rotor
\bea
I_{zz} \gg I_{xx}, \qquad I_{yy} \gg I_{xx}.
\label{moments2}
\eea
With these inequalities, the $L_x$ term in the hamiltonian becomes the dominant part and we denote it by $H_0$. The $L_y$ and $L_z$ terms constitute the perturbative part, and we denote the corresponding summand by $H_{\rm{band}}$. Therefore we have
\bea
H = H_0 + H_{\rm{band}},
\label{zero}
\eea
with the dominant part given by
\bea
H_0 = c \sqrt{\frac{M}{I_{xx}}} \alpha_x L_x + \beta Mc^2
\label{one}
\eea
and a perturbative part
\bea
H_{\rm{band}} = c \sqrt{M} \left( \frac{1}{\sqrt{I_{yy}}} \alpha_y L_y + \frac{1}{\sqrt{I_{zz}}} \alpha_z L_z \right).
\label{two}
\eea
The aforementioned bands comprising the nearly degenerate levels can be obtained from the spectrum of the operator $H_0$: By virtue of the Clifford algebra of the Dirac matrices, we have
\bea
H_0^2 = \frac{Mc^2}{I_{xx}} L_x^2 + M^2 c^4,
\label{oneprime}
\eea
and the position of the bands can be determined by analyzing the spectrum of $L_x$. Interestingly, the details of the spectrum within each band can be described by the perturbative part alone, where again $L_x$ plays an important role. We can show this by computing the square of (\ref{two}) and using the definition $I \equiv I_{zz} = I_{yy}$, leading to
\bea
H_{\rm{band}}^2 = \frac{Mc^2}{I} \left( L^2 - L_{x}^2 - 2 S_x L_x \right).
\label{three}
\eea
At the end of the day, the Zeeman operator $S_x L_x$ is the essential building block of the problem, giving both the position of the band via $H_0$ and the structure of the band via $H_{\rm{band}}^2$. The resulting eigenvalue problem reduces then to the diagonalization of $S_x L_x$ or $L_x$ alone.

Furthermore, our relativistic hamiltonian allows the possibility of taking the ultra relativistic limit $M \rightarrow 0$ with the length parameter $\sqrt{I_{xx}/ M }$ fixed \footnote{The definition of $I_{xx}$ shows it transparently, as the ratio $I_{xx}/M$ tends to $\langle x^2 \rangle$. }. In the extreme case of rods or dumbbells, $H_0$ dominates completely the spectrum and the energy is given directly by $\pm L_{x}$ through the equation
\bea
 c \sqrt{\frac{M}{I_{xx}}} \alpha_x L_x \psi = E \psi.
\label{4.1}
\eea
Once again, we have arrived at the result that $L_x$ alone determines the energy, in this case for an ultra relativistic prolate body. It is important to stress that we have ended up with a finite equispaced spectrum. Such spectra are of more general interest as recently discussed by {'t Hooft} \cite{hoo10}.

\section{Relativistic Rotors and Zeeman interactions: A broader scope}
\label{sec:Zeeman}

We have seen so far that a component of the orbital angular momentum is sufficient to produce the energy location of the rotational bands, as well as the internal structure of the levels within each band. Even the hamiltonian of a massless ultra relativistic rotor could be identified with such a component of $\v L$.

In the context of integrable or exactly solvable models with rotational invariance, it is not an exaggeration to state that the operator along the quantization axis $L_z$ controls everything. However, emulating such a diagonal operator with a homogeneous set of resonators represents a challenge. We have reached a solution of the problem by recognizing that the rotated versions of $L_z$ (namely $L_x$ and $L_y$) are tridiagonal operators whose form is readily implemented in a scheme of nearest neighbour couplings.

In all, one may ask whether more complicated or more general models can be realized using similar schemes. Notably, the answer stems from the nature of the Lie algebra $SO(3)$, which is solvable: Given a irreducible matrix representation of the group $SO(3)$, i.e. given $l$, one can use the Cartan basis of the algebra (or ladder operators $L_{\pm}$) in order to construct multidiagonal operators representing first, second and even multiple neighbour couplings.

Such models are limited in number as the size of the operators is always finite, i.e. $L_{\pm}^{2l + 1} = 0$, which is a direct consequence of the group compactness. In this way one may argue straightforwardly that any hamiltonian $H(L^2,L_z)$ can be represented as a polynomial of $L_x$, making plausible its emulation with our constructions.

It is important to recognize that the previous argumentation is valid even in the presence of spin operators or Dirac matrices $-$ the latter are generators of the Minkwoski Clifford algebra and can be obtained from direct products of spin operators. The emulation of Zeeman terms reduce naturally to $S_z L_z$ or its rotated version $S_x L_x$. The group in question is now $SU(2)_{\rm{spin}} \otimes SO(3)_{\rm{orbit}}$. For a given representation of $SU(2)$ (in our case, $s=\ahalf$), it also holds that a hamiltonian $H(L^2,S_x,L_x)$ is also a polynomial of $L_x$ and that the powers of $H$ eventually eliminate the presence of the spin operators by virtue of the Pauli matrices algebra.

We may look at the following example. The dumbbell kinetic energy operator $K = \sigma_{+}L_{-} + \sigma_{-}L_{+} $ satisfies, upon squaring
\bea
K^2 &=& \sigma_{+}\sigma_{-}L_{-}L_{+} + \sigma_{-}\sigma_{+}L_{+}L_{-} \nonumber \\ &=& \frac{1}{2} \left( 1 + \sigma_3 \right) \left( L^2 - L_z^2-L_z \right) + \frac{1}{2} \left( 1 - \sigma_3 \right) \left( L^2 - L_z^2+L_z \right) \nonumber \\
\fl &=& \left( \begin{array}{cc} L^2 - L_z^2-L_z & 0 \\ 0 & L^2 - L_z^2+L_z \end{array} \right).
\eea
Now the two eigenvalue problems are decoupled and, since $L^2$ is fixed, $L_z$ determines the spectrum of $K$ and it can be emulated by $L_x$.

In conclusion, there are many integrable models in the $SO(3)_{orbital} \times SO(3)_{spin}$ space, but all of them essentially reduce to one tight-binding configuration, which is the emulation of $L_x$ appearing in many fashions either coming from spin or the two signs of the energy in relativistic equations. Most of the arguments we have presented in this section can be applied to any compact Lie algebra, opening the possibility to tight-binding realizations of other physical systems by virtue of a natural map between a Cartan basis and two-point recurrence relations.

\section{Tight-binding realization}
\label{sec:TightBind}

Arrays of identical resonators with nearest neighbour interactions, e.g., potential wells of equal widths and depths, judiciously located, require specific tridiagonal matrices for their realization. A matrix representation of our angular momentum operators can be readily given by fixing the value of $l$ and thereby the dimension of the corresponding Hilbert space. We have seen that an $L_x$ term appears by itself in the hamiltonian describing the position of the respective bands. The eigenvalue problem (\ref{4.1}) can be written conveniently in a spinor basis by recognizing that the presence of $\alpha_x$ only contributes to an overall sign. Using the eigenbasis $\{ D^{l}_{m, m'} \}$ of Wigner rotations for the operators $L^2$ and $L_z$, we obtain

\bea
\fl c \sqrt{\frac{M}{I_{xx}}} \left( \sqrt{(l-m+1)(l+m)} D^{l}_{m-1,m'} + \sqrt{(l-m)(l+m+1)} D^{l}_{m+1,m'} \right) = \pm E D^{l}_{m,m'}. \nonumber \\
\label{4.1.1}
\eea
This equation can be compared with a nearest-neighbour tight-binding relation containing the couplings $\Delta_{m}, \Delta_{m+1}$:
\bea
\Delta_{m} \psi_{m-1} + \Delta_{m+1} \psi_{m+1} + E_0 \psi_m = E \psi_m,
\label{4.2}
\eea
where $E_0$ is the energy of the resonance in an isolated resonator. The required couplings can be read off as

\begin{equation}\label{eq:Deltam}
\Delta_{m} = \epsilon \sqrt{(l-m)(l+m+1)}
\end{equation}
with the convenient definition $\epsilon=c \hbar \sqrt{M/I_{xx}}$, which provides the level spacing. At this point we could consider two possible cases arising from \eref{eq:Deltam}:
semi-integer $l$ (the emulation of half integer spin) and integer $l$ (realization of orbital angular momentum). Although it is the second option what fits our scheme in the realization of a rotor, we shall bear both cases in mind for the rest of the paper.

Concrete realizations of tight-binding arrays demand a specific recipe for the engineering of couplings. For the realization we will use, such couplings will typically depend on the spacing between resonators. In the following section we will present the experimental setup, including the distances $d$ between sites using the functional dependence $\Delta(d)$ to induce a specific level spacing $\epsilon$.

\section{Microwave experiments}
\label{sec:MuExp}

Microwave experiments have been a versatile tool to study questions in non-relativistic quantum mechanics \cite{stoe99}. They have been used in the context of quantum chaos \cite{stoe90,dor90,sri91,alt95a,so95}, scattering theory \cite{alt93b,kuh05a}, spectral statistics \cite{alt94,bar99d,sch01d,sch03a}, disordered systems \cite{cha00,lau07,kuh08a}, fidelity \cite{sch05c,sch05c,hoeh08a}, absorption \cite{men03a,kuh05a} and many others. Recently also the energy spectrum of graphene has been investigated \cite{pel07,bit10a,zan10,bit12,kuh10a,bel13a,bar13a,bel13b}. The hamiltonian of graphene around the so called Dirac points resembles a two dimensional relativistic hamiltonian \cite{sem06}. One microwave realization uses an array of disks with a high index of refraction that are coupled evanescently \cite{kuh10a,bel13a,bar13a,bel13b}. It is an experimental realization of a tight-binding system, which we will adapt now for the realization of the one-dimensional Dirac gyroscope. In the next subsection we describe the setup and adjust the parameters of the set-up in such a way that we minimize effects of higher order neighbour couplings. For details of the relation between the experiment and a tight binding hamiltonian we refer to Ref.~\cite{bar13a,bel13b}.

\subsection{Experimental setup and specifications}
\label{ssec:ExpSetup}

\begin{figure}
\includegraphics[width=.5\columnwidth]{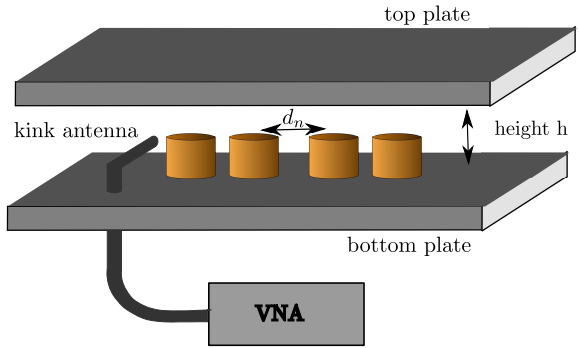}
\caption{\label{fig:Setup}
Sketch of the experimental setup showing the metallic top and bottom plate, the kink antenna and a few disks.}
\end{figure}

The main ingredient of the experiment is a disk with a high index of refraction $n_r\approx$6. The disk has a radius of $r_D$=4\,mm and a height $h_D$=5\,mm. It is sandwiched between two metallic plates which have a distance $h$ between them (see \fref{fig:Setup}). The resonances within the disks are excited using a vector network analyzer connected via a kink antenna to the system. It excites the first TE resonance of the disk \cite{bar13a}. The disks are coupled by evanescent waves as the resonance frequency of the disk is below the cut-off frequency in air, which is induced by the two metallic plates\cite{bel13b}.

\begin{figure}
\includegraphics[width=.8\columnwidth]{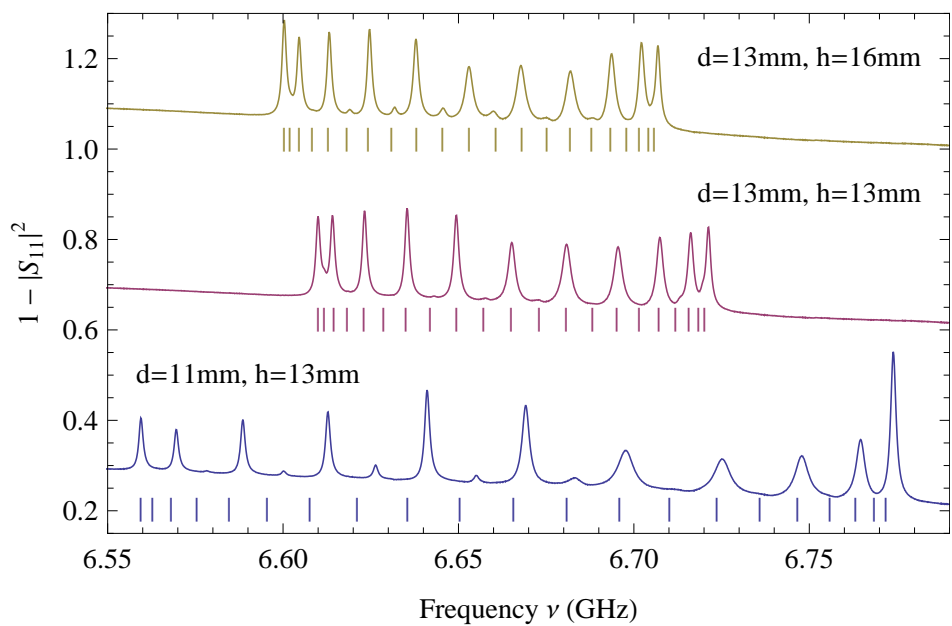}
\caption{\label{fig:EquiDistDisks}
Three experimental reflection spectra for different heights $h$ and distances $d$. The vertical lines below are the corresponding numerical spectra using only nearest neighbour couplings. As the antenna position is at the central disk the even resonances are strongly suppressed. Even though they are hardly visible in the figure, they still can be extracted from the spectra. The parameters of the shown spectra are marked in \fref{fig:Chi2}.}
\end{figure}

\begin{figure}
\includegraphics[width=.8\columnwidth]{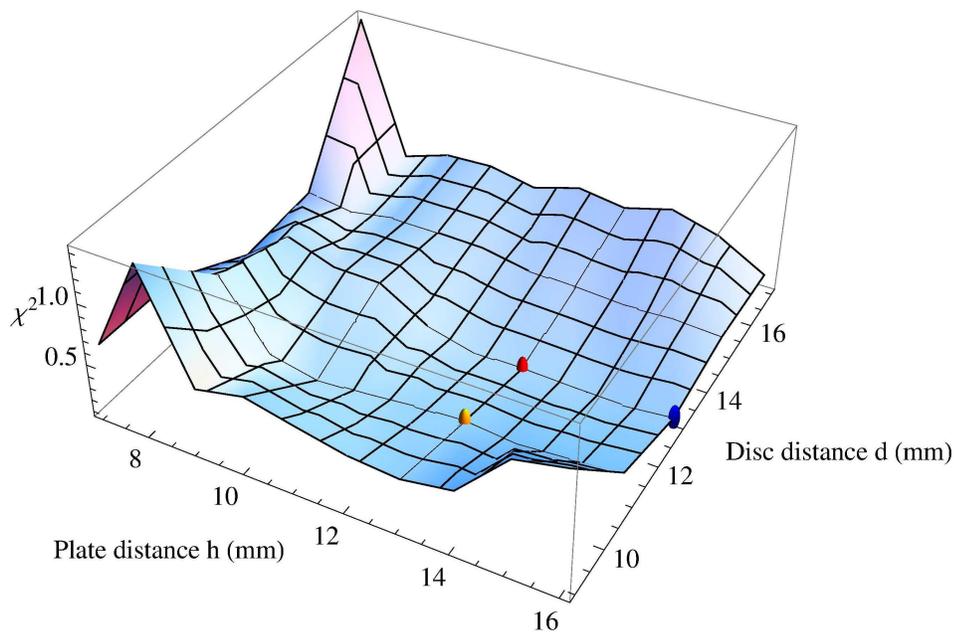}
\caption{\label{fig:Chi2}
The $\chi^2$ value between the experimental resonances compared to the numerical resonances using only nearest neighbour couplings as a function of the plate distance $h$ and the disk distance $d$. The markings refer to the spectra shown in \fref{fig:EquiDistDisks}.}
\end{figure}

To realize the Dirac gyroscope it is necessary that the next-nearest neighbour coupling and all higher order couplings are as small as possible. Thus we first start to adjust the working point. We measured the reflections spectra for 21 equispaced disks for different plate distances $h$ and inter disk distances $d$ (see examples in \fref{fig:EquiDistDisks}). From the measured band width $\delta\nu=\nu_\textrm{max}-\nu_\textrm{min}$ we calculated the next nearest neighbour coupling by $\Delta=\delta\nu/4$ as we have a periodic system. Introducing a tight binding hamiltonian, where the diagonal is set to the eigenfrequency of the disk and the secondary diagonal is set to $\Delta$, we calculated numerically the spectrum. The corresponding eigenfrequencies are indicated by the vertical bars in \fref{fig:EquiDistDisks}. Now we calculated the $\chi^2$ deviation between the experimental and numerical resonance positions. The deviations are presented in \fref{fig:Chi2} as a function of the plate distance $h$ and inter disk distance $d$. We observe a minimal plateau around $h$=12-14\,mm and $d=9-14$\,mm. Thus for all further measurements we fix the plate distance to $h$=13\,mm and will stay within a disk distance of 9-14\,mm. In this range we found that the next-nearest neighbour coupling is less than 7.5\% of the nearest neighbour coupling.

\begin{figure}
\includegraphics[width=.8\columnwidth]{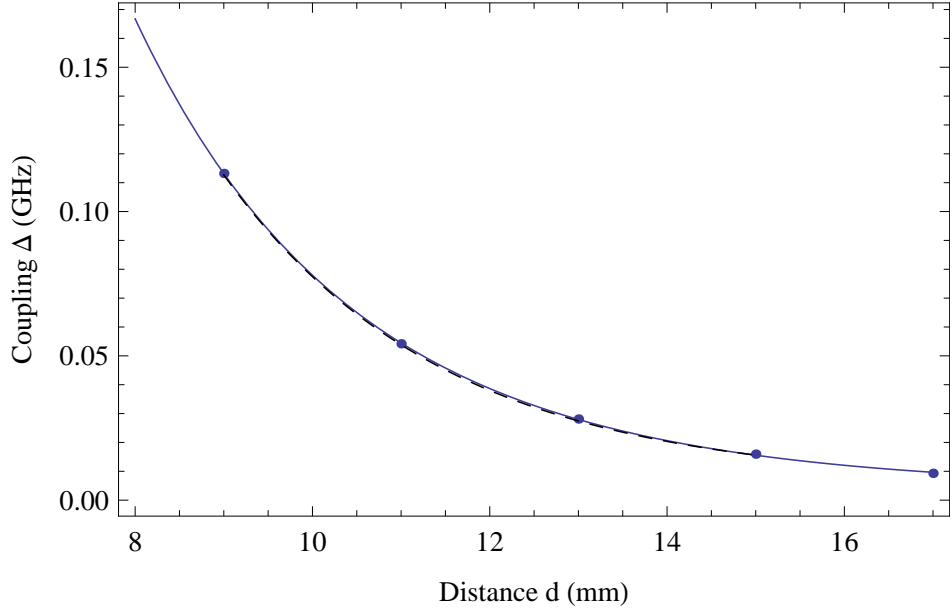}
\caption{\label{fig:Coupling}
Extracted nearest neighbour coupling $\Delta(d)$ of the disks as a function of the disk distance at height $h$=13\,mm. The blue solid line corresponds to a fit to \eref{eq:CouplingK} with $\Delta^\prime_0$=1.53\,GHz, $\gamma^\prime$=0.148\,mm$^{-1}$, and $C^\prime$=0.0039\,GHz and the dashed line to a fit to an approximated exponential \eref{eq:CouplingExp} with $\Delta_0$=3.94\,GHz, $\gamma$=0.20\,mm$^{-1}$ and $C$=0.006 in the range of interest.}
\end{figure}

For $r>r_D$, i.e.~outside a single disk, the eigenfunction is described by a modified Bessel function $K_0$. Thus the coupling between two disks can be estimated by \cite{kuh10a,bar13a}
\begin{equation}\label{eq:CouplingK}
\Delta(d)= \Delta^\prime_0\left|K_0~\left(\gamma^\prime d\right)\right|^2 + C^\prime,
\end{equation}
where $d$ is the center to center distance of the disks. The constant $C^\prime$ takes into account that the resonance frequency of the disks are slightly different. Have in mind that at $d=2r_D$, where $r_D$ is the radius of the disks, the disks are touching. Thus the maximal coupling is given $\Delta(2r_D)$. $\gamma^\prime$ depends strongly on the plate distance $h$. \Fref{fig:Coupling} shows the nearest neighbour couplings $\Delta(d)$ extracted via the band width as a function of the inter disk distance. The blue line corresponds to a fit of \eref{eq:CouplingK}. In the range of interest $d$=9-14\,mm the coupling can be also approximated by an exponential
\begin{equation}\label{eq:CouplingExp}
\Delta(d) = \Delta_0 \exp \left(-\gamma d\right) + C,
\end{equation}
which is indicated by the dashed line in \fref{fig:Coupling}. Now we have adjusted the working point and obtained all necessary ingredients to setup the experiment for the Dirac gyroscope.

\subsection{Spectrum of a Dirac gyroscope}
\label{ssec:ExpSpecGyro}

As we have shown in \sref{sec:TightBind} we need to adjust the couplings $\Delta_n$ corresponding to \eref{eq:Deltam}. For the sake of simplicity we use here the exponential description of the coupling. Thus we invert first \eref{eq:CouplingExp} resulting in
\begin{equation}\label{eq:DistanceDelta}
d(\Delta) = -\frac{1}{\gamma} \ln \left(\frac{\Delta}{\Delta_0}\right).
\end{equation}
Taking into account the relation \eref{eq:Deltam} for the couplings we get the relation for the disk distances
\begin{equation}\label{eq:Distancem}
d_n = d_{l+m+1} = -\frac{1}{\gamma}\ln \left(\frac{\epsilon}{\Delta_0}\sqrt{(l-m)(l+m+1)} \right),
\end{equation}
where $N=2l+1$ and $m=n-l-1$. In the experiment $\epsilon$ is given in GHz and is defining the level spacing. Have in mind that $n$ is the consecutive disk number ranging from 1 to $N=2l+1$, which is the total number of disks, whereas in \eref{eq:Deltam} we have $l$ and $m$ ranging from $-l$ to $l$. The system is symmetric with respect to the center, where the largest distance is at the center and the minimal distance at the edge is defining the level spacing by
\begin{equation}\label{eq:Epsilon}
\epsilon=\frac{\sqrt{l(l+1)}}{\Delta_0}\exp\left(\gamma d_\textrm{min}\right).
\end{equation}
The consecutive difference is given by
\begin{equation}\label{eq:deltadm}
\delta d_m=d_{m+1}-d_m=-\frac{1}{\gamma}\log\left(\frac{\sqrt{(l-m+1)(l+m+2)}}{\sqrt(l-m)(l+m+1)}\right)
\end{equation}
It is important to mention that in experiments involving microwaves we must introduce additional constants $E_0$, which is a global shift that takes into account the resonance frequency of the single disk. In the following we shall fix these quantities according to our setup.

\begin{figure}
\includegraphics[width=.8\columnwidth]{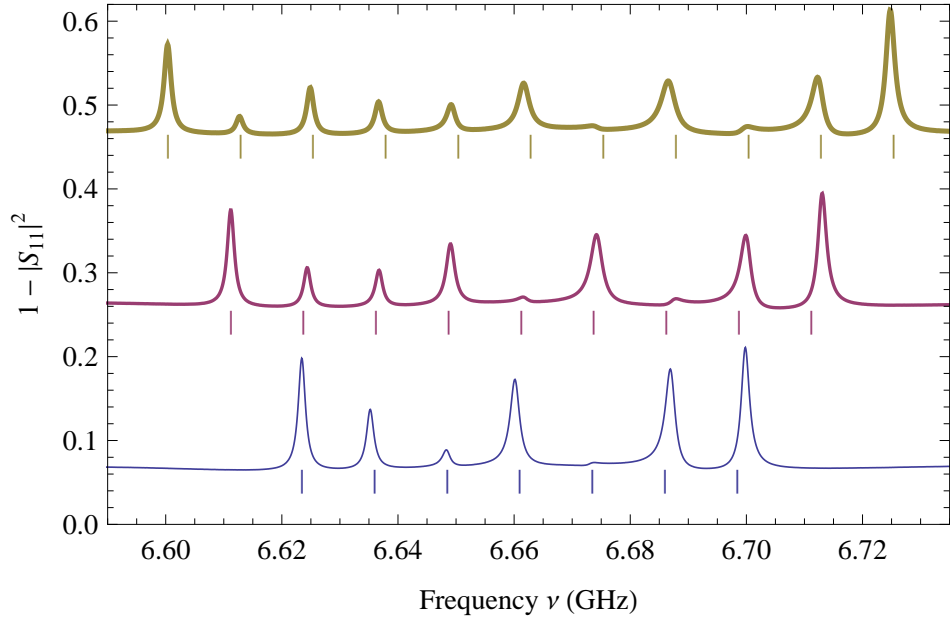}
\caption{\label{fig:GyroscopeRSpectra}
Experimental reflection spectra $1-|S_{11}|^2$ for three realizations of the gyroscope with different number of disks and the same level spacing $\epsilon = 25 {\rm MHz}$. The bars indicate the eigenvalues of the corresponding prediction from \eref{eq:Epsilon}, where the central energy is fixed by means of the lowest experimentally obtained resonance. The spectra are shifted in increasing order and correspond to 7, 9 and 11 disks.}
\end{figure}

We now want to realize a Dirac gyroscope with fixed level spacing for different number of disks $N$. We chose $\epsilon$=25\,MHz and realized three chains with $N$=7,\ 9,\ 11. Using \eref{eq:Distancem} we can calculate the corresponding disk distances $d_n$ necessary to set up the different chains of disks. In \fref{fig:GyroscopeRSpectra} we show the three corresponding measured reflection spectra. The bars indicate the expected equidistant Dirac gyroscope spectra, where the central energy is fixed by means of the lowest experimentally obtained resonance. This procedure takes also into account an additional shift induced by higher order couplings. Good agreement between the reflection spectra and theory is found.

\begin{figure}
\includegraphics[width=.8\columnwidth]{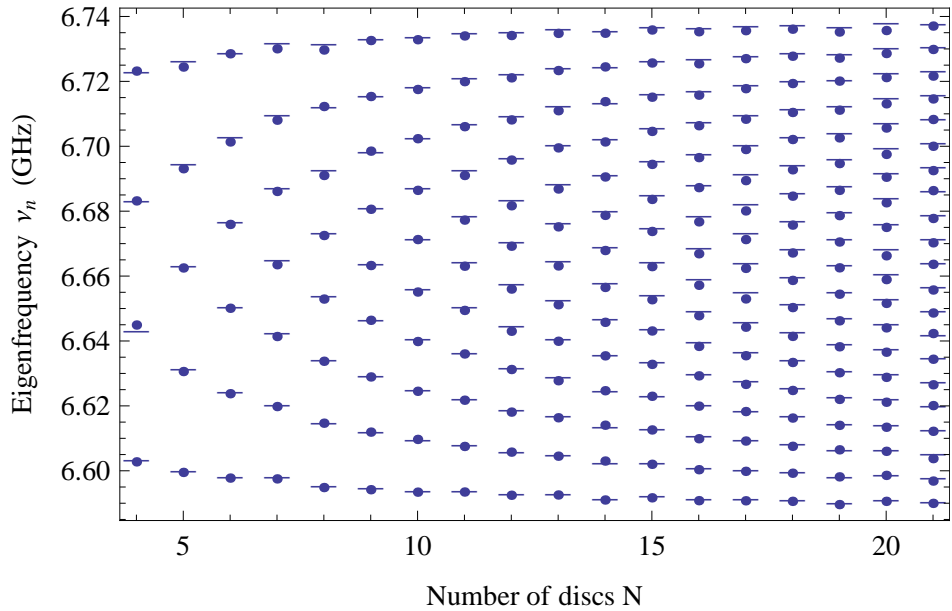}
\caption{\label{fig:GyroscopeNChange}
Eigenfrequencies $\nu_n$ for different numbers of disks with $d=13$\,mm.
The dots are the experimental eigenfrequencies obtained by the maxima of $1-|S_{11}|^2$ and the lines correspond to the theoretical prediction, where the lowest frequencies were adjusted.}
\end{figure}

\begin{figure}
\includegraphics[width=.8\columnwidth]{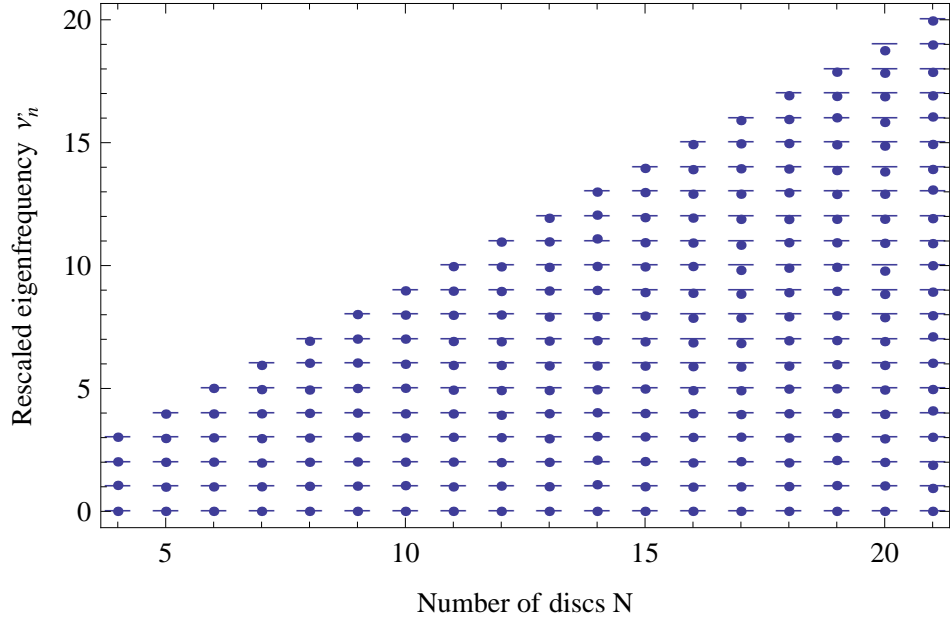}
\caption{\label{fig:GyroscopeNChangeRescaled}
Rescaled eigenfrequencies $\nu^\prime_n$ for different numbers of disks with $d=13$\,mm.}
\end{figure}

Next we fixed the minimal length $d_\textrm{max}$=13\,mm for $N$=4 disks. We added step by step an additional disk corresponding to \eref{eq:deltadm}. We did this up to $N$=21. In \fref{fig:GyroscopeNChange} the corresponding experimental resonances are shown as dots, whereas the horizontal bars indicate the levels for the equidistant spectra. The experimental spectra show the expected change of the level spacing, which is due to the fact that, with each additional disk, the minimal distance is reduced, thus also leading to a reduction of the level spacing (see \eref{eq:Epsilon}). Finally we rescale the resonance by
\begin{equation}\label{eq:Rescaling}
\nu'_n= (\nu_n-\nu_0)/\Delta\nu,
\end{equation}
where $\Delta\nu$ corresponds to the theoretically predicted spacing and $\nu_0$ is the first resonance extracted from the experimental spectra for each number of disks, respectively. Thus $\nu'_n$ should resemble the integer level number if the spectra are evenly spaced. This is shown in fig.~\ref{fig:GyroscopeNChangeRescaled} and the predicted behaviour is found.

The results show that it is possible to engineer a system with a complete set of quantum numbers using tight-binding arrays. The general spectroscopic structure of the gyroscope has been therefore reproduced: The columns of fixed angular momentum in \fref{fig:GyroscopeNChangeRescaled} accommodate an increasing number of states, representing multiplets of such an observable. With our results we establish a connection with the general trend of the different levels of the axi-symmetric case presented in figure \ref{fig3}. Furthermore, the baryonic spectra presented also in figure \ref{fig3} possess an increasing number of states which can be understood as splittings for low masses, opening the possibility of emulating nearly any bound spectrum of fixed $l$ through deformations of our previous construction.

It is important to note that the phenomenology of the mass states of baryons becomes more involved as the energy increases. The corresponding resonances might be accommodated in new rotational bands, but in such a case any model of rigidity (either from strict or loosened requirements for relativistic systems) should be abandoned, as possible vibrational states should dominate the picture.

In all, our example of an equispaced spectrum can be regarded as a benchmark for more detailed constructions that adjust levels for each observable (e.g. the value of $l$). The experimental results in \fref{fig:GyroscopeNChangeRescaled} reproduce such a physical situation. Interestingly, techniques of this type resemble those of chemistry in the form of polyads \cite{jun02,her13}: building the skeleton of the spectrum and later enrich it and/or perturb it has proven to be a sensible approach.

\section{Discussion and outlook}
\label{sec:DiscOutlook}

Taking up a suggestion made by one of the authors in a previous paper \cite{sad09} to use a relativistic rotor or gyroscope as a schematic model for baryons, we have emulated such a system in a microwave setup after mapping it onto a chain of resonators consistent of dimers with successively decreasing coupling. This was implemented by increasing distances between resonators taking advantage of the fact that the selected TE mode was evanescent between the resonators for the selected distance of the covering plate. The fact that we have finite spectra for the relativistic rotor eliminates a priori an important source of errors in the case of infinite spectra namely the inevitable truncation. We have thus achieved an emulation of a relativistic system of relevance and the agreement between experiment and theory is satisfactory.

Wave functions can be analyzed, but the fact that the eigenfunctions are essentially Jacobi polynomials suggests that the rotary structure will be visible when explored in future work. The model relies on the fact that there are nearest neighbour interactions only, while our experimental setup makes it difficult to suppress higher order neighbour interactions, if we wish to use a two-dimensional array. Yet considering that only the coupling strength and the topology determine such a model, we propose to use quantum graphs, e.g. coupling by cables or wave guides of sets of equal resonators that have an isolated resonance in the frequency domain we wish to study. For systems where no cutoff is needed we can expect that the quality of the resonators will determine the quality of the emulation. As to the wavefunctions, it is possible to retrieve them experimentally by measuring the height of the peaks as a function of the disc number. Summarizing we have a promising field to emulate a wide variety of relativistic and non-relativistic systems by more schematical or more realistic schemes.

\ack

We are thankful to S. Barkhofen for fruitful discussions.

T. H. S. and J. A. F. V. thank the LPMC for the hospitality during several long term visits and to CONACyT Project Number 44020 and PAPIIT-UNAM project number IG101113 for financial support. E. S. acknowledges support from PROMEP project No. 103.5/12/4367.

\section*{References}
\label{sec:references}

\end{document}